\documentclass[a4paper,fleqn]{cas-dc}

\usepackage[authoryear]{natbib}
\def\tsc#1{\csdef{#1}{\textsc{\lowercase{#1}}\xspace}}
\tsc{WGM}
\tsc{QE}
\tsc{EP}
\tsc{PMS}
\tsc{BEC}
\tsc{DE}
\begin{document}
\let\WriteBookmarks\relax
\def\floatpagepagefraction{1}
\def\textpagefraction{.001}
\shorttitle{Nonlinear acoustic-gravity waves}
\shortauthors{D. Chatterjee et~al.}
\title [mode = title]{ Effects of Coriolis force on the nonlinear interactions of acoustic-gravity waves in the atmosphere}                      
%\tnotemark[1]
%\tnotetext[1]{This document is the results of the research
   %project funded by the National Science Foundation.}

%\tnotetext[2]{The second title footnote which is a longer text matter
  % to fill through the whole text width and overflow into
   %another line in the footnotes area of the first page.}

%\author[1,3]{CV Radhakrishnan}[type=editor,
%                        auid=000,bioid=1,
%                        prefix=Sir,
%                        role=Researcher,
%                        orcid=0000-0001-7511-2910]
\author[1]{Debjani Chatterjee} 
%\fnmark[1]
\ead{chatterjee.debjani10@gmail.com }
%\ead[URL]{ }
\address[1]{ Department of Applied Mathematics, University of Calcutta,  Kolkata-700 009, India}
%%%%%%%%%
\author[2]{A.P. Misra}[orcid=0000-0002-6167-8136]
\cormark[2]
%\fnmark[2]
\ead{apmisra@visva-bharati.ac.in; apmisra@gmail.com}
%\ead[url]{www.cvr.cc, cvr@sayahna.org}
%\credit{Conceptualization of this study, Methodology, Software}
\address[2]{ Department of Mathematics, Siksha Bhavana, Visva-Bharati (A Central University), Santiniketan-731 235, India}
\cortext[cor2]{Corresponding author}
%\cortext[cor2]{Principal corresponding author}
%\fntext[fn2]{ }
%\fntext[fn2]{Another author footnote, this is a very long footnote and
  %it should be a really long footnote. But this footnote is not yet
  %sufficiently long enough to make two lines of footnote text.}

%\nonumnote{This note has no numbers. In this work we demonstrate $a_b$
%  the formation Y\_1 of a new type of polariton on the interface
%  between a cuprous oxide slab and a polystyrene micro-sphere placed
%  on the slab.
%  }

\begin{abstract}
 The nonlinear theory of acoustic-gravity waves (AGWs) in the atmosphere is revisited with the effects of the Coriolis force.  Previous theory in the literature [Phys. Scr. \textbf{90} (2015) 055001] is advanced.   Starting from a set of fluid equations modified by the Coriolis force, a general linear dispersion relation is derived which manifests the  coupling of the high-frequency acoustic-gravity waves (AGWs) and the low-frequency  internal gravity waves (IGWs). 
The frequency of IGWs is  enhanced by the Earth's angular velocity.   The latter also significantly modifies the  nonlinear coupling of AGWs and IGWs   whose evolutions are described by   the Zakharov approach as well as the  Wigner-Moyal formalism.  The consequences of AGWs and the two equivalent evolution equations modified by the Coriolis force are briefly discussed. 

\end{abstract}
 
%%%%%%%%%%%%%%%%%%%%%%%%%%%%%%%%%%%%%%%%%%%%%%%%%%%%%%%%%%%%%%%%%%%%%
%\begin{highlights}
%\item The effect of Coriolis force on the nonlinear propagation of acoustic-gravity waves (AGWs)  is studied 
%\item The nonlinear coupling of high- and low-frequency  AGWs can be described by either the Zakharov approach or the Wave-kinetic approach
%\item The  high-frequency wave ponderomotive nonlinearity is enhanced and the resonance condition is modified by the  Coriolis force  
%\item Both the Zakharov and the  wave-kinetic approach are useful for the description of atmospheric turbulence
%\end{highlights}
%%

\maketitle

\section{Introduction}\label{sec-intro}
   The propagation of acoustic-gravity waves (AGWs) has been known to   play a significant role in the interpretation of a  wide varity of wave phenomena in the atmosphere including those in the troposphere, as well as to describe the dynamics of  ionospheric plasmas \citep{hines1960,hook1968}. The atmospheric waves, whose frequency is of the order of the Brunt-V{\"a}is{\"a}l{\"a} frequency or   buoyancy frequency and for which  the potential energy associated with the buoyancy frequency becomes almost equal to the kinetic energy plus the elastic energy of the acoustics, are termed as AGWs. The frequency of the latter  is much lower than that human ears can detect it as sound waves. However,  they have  some visible impacts in the patterns of atmospheric clouds. Furthermore, the AGWs can be useful for predicting   weather and climate phenomena   for detecting and monitoring the nuclear detonations as well as    to describe the dynamics of the global atmospheric turbulence. The importance of such AGWs has been recognized by a number of authors  in the linear and nonlinear regimes of lower and upper atmospheres, as well as in the Earth's E- and F-layers \citep{stenflo1987,stenflo1998,stenflo2009,kaladze2007,kaladze2008,mendonca2014,roy2019}. It has been investigated that the AGWs can also appear as a consequence of various meterological and auroral conditions  including the  solar eclipses and earthquakes of shear flows \citep{jovanovic2002}. Other important consequences of  the AGWs are the formation of localized solitary structures, solitary vortices \citep{kaladze2008} and the onset of turbulence due to the interactions of  high- and low-frequency branches of AGWs \citep{mendonca2015}.  
   \par
Various appealing  phenomena  occur  when the Coriolis force due to the Earth's rotation  with the angular velocity $\bf{\Omega}_0$ is considered in the fluid dynamics. The Coriolis force not only gives rise to the coupling of high- and low-frequency AGWs but also modifies the resonance and cut-off frequencies of various other modes  in the atmosphere. It has been shown that such a force in incompressible fluids can also lead to the evolution of solitary vortices \citep{kaladze2008}.  
\par Because of the existence of two frequency branches of AGWs, namely the high-frequency AGWs and the low-frequency internal gravity waves (IGWs),  various nonlinear theories of wave-wave interactions have been explored in the context of Zakharov approach  \citep{stenflo1986,mendonca2015}. The latter is not only useful for the evolition of solitons associated with the high-frequency wave fields but also for the description of chaos and  fluid turbulence by the process of energy transfer in nonlinear media.  An alternative approach, namely the wave-kinetic approach based on the Wigner-Moyal formalism, has also received considerable attention  for the description of the nonlinear coupling of   high- and low-frequency branches of AGWs \citep{mendonca2014,mendonca2015,mendonca2006}.  Such an approach, based on two-fluid model,  was first proposed by Tisza, and executed by Landau \citep{leggett2006}.  Later, this approach has been adapted in several fields including the atmospheric physics and the plasma physics \citep{mendonca2006a}.  Recently, Mendonca and Stenflo \citep{mendonca2015} developed the wave-kinetic theory of AGWs in the atmosphere  starting from a set of Zakharov-like equations \citep{zakharov1972} without the influence of the Coriolis force. They remarked that  the Zakharov and wave-kinetic approaches are nearly equivalent and they   can provide two complementary views of the  atmospheric  turbulence.
\par
In this work, our aim is to revisit the nonlinear theory of AGWs, especially to advance the work of Mendonca and Stenflo \citep{mendonca2015} with the influence of the Coriolis force in the fluid motion of charged particles. Starting from a set of nonlinear fluid equations for AGWs,   we derive  a set of modified Zakharov-type equations which govern  the nonlinear interactions  of two different frequency branches of AGWs. Based on the Wigner-Moyal formalism, we also derive an equivalent coupled wave-kinetic equations that are modified by the Coriolis force.  We find that  the ponderomotive nonlinearity is enhanced and the Landau resonant velocity is up-shifted by the effects of the Coriolis force.  
   %%%%%%%%%%%%%%%%%%%%%%%%%%%%%% 
 \section{Theoretical Formulation} \label{sec-formul}
We consider the nonlinear propagation of AGWs in a weakly ionized  atmospheric conducting fluid with density $\rho$, the pressure $p$ and the velocity ${\bf v}$.     We assume that the Coriolis force on the charged particles is due to the Earth's rotation with the uniform angular velocity ${\bf{\Omega}}_0\equiv(0,0,\Omega_0)$ along the vertical direction.    It is further assumed that the  atmospheric conducting fluid is  unmagnetized for which there is no  influence of the Amp{\'e}re force. Such an assumption may be valid in the lower region of the Earth's atmosphere, e.g., inospheric D-region  \citep{kaladze2008}.  Also, we assume that the conducting fluid is  quasi-neutral for which  the inner electrostatic electric field can be neglected, i.e., ${\bf{E}}=-\nabla\phi={\bf 0}$.  Here, $\phi$ is the electrostatic potential.  Thus, the dynamics of atmospheric fluids can be described by the following sets of  equations.
\begin{equation}
\frac{\partial \rho}{\partial t}+ \nabla \cdot (\rho {\bf{v}})=0, \label{continuity-eq}
\end{equation}
\begin{equation}
\frac{\partial \bf{v}}{\partial t}+ {\bf{v} \cdot \nabla \bf{v}} =-\frac{\nabla p}{\rho} -2 {\bf{\Omega}}_0 \times \bf{v} +\bf{g}, \label{momentum-eq}
\end{equation}
\begin{equation}
\left(\frac{\partial }{\partial t}+ {\bf{v \cdot \nabla}}\right)  (\rho^{-\gamma} p)  =0, \label{pressure-eq}
\end{equation}
 where  ${\bf{g}}=(0,0,-g)$ is the gravitational acceleration and $\gamma$ is the ratio of the specific heats. 
 At equilibrium,  the background mass density and  pressure  can be assumed to vary as $\rho_0(z)=\rho_0(0)\exp(-z/H)$ and $p_0(z)=p_0(0)\exp(-z/H)$,  where $p_0~(\rho_0)$ is the background pressure (mass density) stratified by the gravitational field  and $H$ is the  reduced scale length of the   atmosphere, i.e.,  $H=c_s^2/\gamma g$ with $c_s$ denoting the sound speed.  
\par 
In what follows, we linearize Eqs. \eqref{continuity-eq} - \eqref{pressure-eq} by splitting up the physical quantities into their equilibrium (with suffix $0$) and perturbation (with suffix $1$) parts. Introducing a new variable   $N\equiv \rho_1/ \sqrt{\rho_0}$  with $\rho_1=\rho-\rho_0$, and following Ref. \citep{stenflo1986} we obtain the following modified evolution equation for the density perturbation of  AGWs.
\begin{equation}
\begin{split}
&\sqrt{\rho_0}\left[\frac{\partial^4}{\partial t^4} +\left( \omega_a^2-c_s^2 {\bf{\nabla}}^2+4\Omega_0^2\right)  \frac{\partial^2}{\partial t^2} \right. \\ 
 &\left.-c_s^2\omega_g^2 \nabla_\perp^2+4\Omega_0^2 \omega_a^2-4c_s^2\Omega_0^2 \frac{\partial^2}{\partial z^2}\right]N  =0, \label{linear-eq} 
\end{split}
\end{equation}
where $\nabla_\perp^2 =\partial^2/\partial x^2+\partial^2/\partial y^2$, and  $\omega_a$ and $\omega_g$ are two characteristic frequencies, given by, $\omega_a^2\equiv c_s^2/4H^2$ and $\omega_g^2 \equiv (\gamma-1)c_s^2/\gamma^2 H^2$. In fact, these two frequencies  define two distinct wave modes to be obtained shortly. Next, we  derive the linear dispersion relation from Eq. \eqref{linear-eq} by assuming  the density perturbations  to vary as plane waves with frequency $\omega$ and the wave vector ${\bf k}$, i.e.,   $N\propto\exp(i{\bf{k}\cdot\bf{r}}-i\omega t)$. Thus, we obtain   \citep{kaladze2008}
\begin{equation}
\begin{split}
\omega^4&-\omega^2\left(\omega_a^2+k^2c_s^2+4\Omega_0^2\right)\\
&+c_s^2\omega_g^2 k_\perp^2+4\Omega_0^2(c_s^2k_z^2+\omega_a^2)=0.
\end{split}
\label{dispersion-eq}
\end{equation} 
The dispersion equation \eqref{dispersion-eq} agrees with that obtained by Kaladze \textit{et al.} for AGWs \cite{kaladze2008}.  From Eq. \eqref{dispersion-eq} it is noted that the dispersion of AGWs is significantly modified by the Coriolis force $(\propto\Omega_0)$. In fact,   Eq. \eqref{dispersion-eq} gives two wave modes in two different frequency limits. In the limit of $\omega \gg\omega_a~(>\omega_g>\Omega_0)$ we obtain the high-frequency (with subscript $h$) acoustic-gravity mode, given by,
\begin{equation}
\omega_h^2=\omega_a^2+k^2c_s^2,\label{dispersion-high-freq}
\end{equation} 
while in the opposite limit, i.e., $\omega\ll\omega_g$, the low-frequency (with subscript $l$) internal gravity mode is obtained, i.e.,   
\begin{equation}
\omega_l^2= \frac{k_\perp^2 \omega_g^2}{k_z^2+{1}/{4H^2}} +\omega_i^2. \label{dispersion-low-freq}
\end{equation} 
Here,  $\omega_a$ and $\omega_i=2\Omega_0$, respectively, represent the cut-off frequencies corresponding to the high-frequency acoustic mode and  the low-frequency internal wave. We note that while the high-frequency mode remains unaltered, the frequency of the internal wave mode and hence its phase velocity  are increased by the effect of $\Omega_0$.   However, these two modes can be nonlinearly coupled.  In the following two sections \ref{sec-zakha} and \ref{sec-wavekin}, it will be shown  that the Coriolis force significantly modifies the nonlinear coupling of the AGWs and IGWs.
 %%%%%%%%%%%%%%%%%%%%%%%%%%%%%%%%% 
\section{Nonlinear evolution equations} \label{sec-zakha}
From Eqs. \eqref{continuity-eq}-\eqref{pressure-eq} and following Refs. \citep{stenflo1986,mendonca2015},   the  evolution equations  for the high-frequency $(N_h)$ and low-frequency $(N_l)$ perturbations  are obtained as 

\begin{equation}
\begin{split}
&\sqrt{\rho_0}\left(\frac{\partial^2}{\partial t^2}+\omega_a^2-c_s^2 {\bf{\nabla}}^2 \right) N_h  \\
 &=\nabla \cdot \left(\rho_0 {\bf{v}}_h\cdot {\bf{\nabla v}}_l+ \rho_0 {\bf{v}}_l\cdot {\bf{\nabla v}}_h -\sqrt{\rho_0} {\bf{v}}_l\frac{\partial N_h}{\partial t}\right), \label{high-freq-eq} 
\end{split}
\end{equation}
%%%%%%%%%%
\begin{equation}
\begin{split}
&\sqrt{\rho_0}\left( \frac{1}{4H^2}\frac{\partial^2}{\partial t^2} -\nabla^2 \frac{\partial^2}{\partial t^2} -\omega_g^2 \nabla_\perp^2\right. \\
&\left.-4\Omega_0^2\frac{\partial^2}{\partial z^2}+\frac{\Omega
_0^2}{H^2} \right)N_l = S(N_h,{\bf{v}}_h), \label{low-freq-eq} 
\end{split}
\end{equation}
where $S(N_h,{\bf{v}}_h)\approx <S(\rho_{1h},{\bf{v}}_h,p_{1h})>\approx c_s^2S$ and $c_s^2 S$ is given by
\begin{equation}
\begin{split}
&c_s^2S\approx -\frac{\sqrt{\rho_0}}{2}\left( \frac{1}{4H^2}\frac{\partial^2}{\partial t^2} -\nabla^2 \frac{\partial^2}{\partial t^2} -\omega_g^2 \nabla_\perp^2\right.\\
&\left.-4\Omega_0^2\frac{\partial^2}{\partial z^2}+\frac{\Omega
_0^2}{H^2} \right) \left\langle \sqrt{\rho_0}\left( {\bf{v}}_h^2- \frac{c_s^2 \rho_{1h}^2}{\rho_0^2}\right)  \right\rangle. 
\end{split}
\end{equation}
\par
Next, introducing a new function $\phi= \int^t \int^z dt dz p_1$, the variables   $N_h,~ {\bf v}_h,~ N_l$ and ${\bf v}_l$ are expressed as
\begin{equation}
N_l=-\frac{H}{\sqrt{\rho_0} c_s^2\left(1+ {16\Omega_0^2H^2}/{c_s^2}\right) } \nabla^2 \left(\frac{\partial \phi_l}{\partial t}\right), \label{N1-eq}
\end{equation}
\begin{equation}
{\bf{v}}_l=-\frac{1}{\rho_0}\left(\hat{z}\nabla^2 \phi_l -\nabla \frac{\partial \phi_l}{\partial z} \right), \label{v1-eq} 
\end{equation}
\begin{equation}
N_h=\frac{1}{c_s^2 \sqrt{\rho_0}}\frac{\partial}{\partial t}\frac{\partial \phi_h}{\partial z}, \label{Nh-eq}
\end{equation}
\begin{equation}
{\bf{v}}_h=-\nabla\left(\frac{1}{\rho_0} \frac{\partial \phi_h}{\partial z}\right)  -2 \frac{H^2}{c_s^2} {\bf{\Omega}}_0\times \nabla \left( \frac{1}{\rho_0}\frac{\partial}{\partial t}\frac{\partial \phi_h}{\partial z}\right). \label{vh-eq}
\end{equation}
\par
To simplify the formalism further, we again introduce the high-frequency variable   $M_h\equiv \int^t N_h(t_1)dt_1$ and the low-frequency variable associated with  the current, i.e.,  ${\bf{j}}_l\equiv \rho_0 {\bf{v}}_l$. Thus,   Eqs. \eqref{high-freq-eq} and \eqref{low-freq-eq} reduce to
\begin{equation}
\begin{split}
&\sqrt{\rho_0}\left(\frac{\partial^2}{\partial t^2}+\omega_a^2-c_s^2 {\bf{\nabla}}^2 \right) \frac{\partial M_h}{\partial t} \\
& =\nabla \cdot \left[ {\bf{v}}_h\cdot {\bf{\nabla j}}_l+ {\bf{j}}_l\nabla\cdot {\bf{v}}_h +{\bf{j}}_l\cdot \nabla {\bf{v}}_h\right], \label{zakharov-eq-1} 
\end{split}
\end{equation}
\begin{equation}
\begin{split}
\frac{\partial {\bf{j}}_l}{\partial t}=&-\frac{c_s^4}{4H}\left( 1+\frac{16\Omega_0^2H^2}{c_s^2}\right) \left(1+\frac{2\Omega_0H}{c_s} \right)\\
&\times \left[\nabla_\perp \frac{\partial}{\partial z}- \hat{z}\nabla_\perp^2  \right]<M_h^2>, \label{zakharov-eq-2} 
\end{split}
\end{equation}
with
\begin{equation}
{\bf{v}}_h=-c_s^2 \nabla\frac{M_h}{\sqrt{\rho_0}}-2 H^2 {\bf{\Omega}}_0\times \nabla \left( \frac{1}{\sqrt{\rho_0}} \frac{\partial M_h}{\partial t}  \right). \label{final-vh-eq} 
\end{equation}
Equations \eqref{zakharov-eq-1} and \eqref{zakharov-eq-2} are the desired Zakharov-like equations for the description of the nonlinear interactions of high-frequency AGWs and the low-frequency current density perturbations of IGWs that are driven by the ponderomotive force of the high-frequency density perturbations.   The appearance of the new terms $\propto\Omega_0$ in ${\bf v}_h$ indicates that the velocity component of the high-frequency field is enhanced by the influence of the Coriolis force. Furthermore, the latter not only modifies the local nonlinear coupling but also significantly enhances the ponderomotive nonlinearity.  Such an enhancement may lead to an increase of the soliton amplitude to be formed in the coherence state as well as may favor the   intermediate chaotic processes  to develop  faster than that in absence of the Coriolis force. We note that in absence of the effects of the Coriolis force, Eqs. \eqref{zakharov-eq-1} and \eqref{zakharov-eq-2} exactly agree with those in Ref. \citep{mendonca2015}. In the next section \ref{sec-wavekin}, we derive an equivalent set of wave-kinetic equations for the nonlinear coupling of AGWs and IGWs, and show that the Landau resonance condition is also modified by the Coriolis force. 
\section{Wave-kinetic equations} \label{sec-wavekin}
We derive the wave-kinetic equations from Eqs. \eqref{zakharov-eq-1} and \eqref{zakharov-eq-2}. Here,  we describe the nonlinear coupling of the high- and low-frequency waves   not in terms of the field amplitudes, but in terms of a quasi-probability where  the high-frequency waves   are described in terms of quasi-particles.  Thus, the high-frequency perturbations can be described by a superposition of plane wave modes with amplitude $M_k$, given by,
\begin{equation}
M_h({\bf{r}}, t)=\int M_k \exp(i{\bf{k}}\cdot{\bf{r}}-i\omega t)\frac{d\bf{k}}{(2\pi)^3},\label{M-h eq}
\end{equation}
where the wave frequency $\omega$ and the wave vector ${\bf k}$ are related to the nonlinear dispersion relation, to be obtained from Eq. \eqref{zakharov-eq-1},   as
\begin{equation}
\begin{split}
\left[\omega^2-(\omega_a^2+c_s^2k^2)\right]& M_k= \frac{{\bf{k}}}{\omega \sqrt{\rho_0}} \cdot \left[{\bf{v}}_k \cdot \nabla {\bf{j}}_l \right.\\
&\left.+ {\bf{j}}_l(i{\bf{k}} \cdot {\bf{v}}_k)+ i({\bf{j}}_l \cdot {\bf{k}}) {\bf{v}}_k\right], \label{high-freq-dispersion-new}
\end{split}
\end{equation}
where
\begin{equation}
{\bf{v}}_k={{\bf{\tilde v}}_k} -i\frac{2H^2\omega}{c_s^2} {\bf{\Omega}}_0 \times {{\bf{\tilde v}}_k}, \label{v-k-eq}
\end{equation}
with
\begin{equation}
{{\bf{\tilde v}}_k}=-\frac{c^2}{\sqrt{\rho_0}}(i{\bf{k}}+k_0{\bf{e}}_z)M_k, \label{v-k1-eq}
\end{equation}
and $k_0=1/2H$ is related to the scale length $H$ of the atmosphere. 
\par 
Next, using Eqs. \eqref{v-k-eq} and \eqref{v-k1-eq},  Eq. \eqref{high-freq-dispersion-new} can be rewritten as
\begin{equation}
\omega^2-(\omega_a^2+c_s^2k^2)= -\frac{c_s^2}{\rho_0} {\cal L}_k {\bf{j}}_1({\bf{r}},t), \label{dispersion-relation}
\end{equation}
where the nonlinear coupling operator ${\cal L}_k$ is given by
\begin{equation}
\begin{split}
&{\cal L}_k=\frac{\bf k}{\omega}\cdot \left[\left\lbrace (i{\bf k}+k_0{\bf e}_z)\right.\right.\\
&\left.\left.-i\frac{2H^2\omega}{c_s^2} \left\lbrace {\bf\Omega}_0 \times (i{\bf k}+k_0{\bf e}_z)\right\rbrace \right\rbrace  \cdot (i{\bf k}+\nabla) \right.\\
&\left.+(ik_0 k_z-k^2)(1-i\frac{2H^2\omega}{c_s^2}\Omega_0)\right]. \label{nonlinear-coupling-operator}
\end{split}
\end{equation}
Equation \eqref{dispersion-relation} can be stated as the local nonlinear dispersion relation of AGWs with the local nonlinear coupling $\propto {\cal L}_k  j_l$ being associated with the slowly varying current density of IGWs. Note that this nonlinear coupling is significantly  modified by the Coriolis force and without which    Eq. \eqref{dispersion-relation} recovers the linear dispersion relation for the high-frequency branch \eqref{dispersion-high-freq}.    We note that  this nonlinear coupling not only modifies  the linear dispersion relation but also introduces a number of new effects including those lead to the collision and fusion among solitons to take place and the emergence of spatio-temporal chaos due to irregular interactions of high- and low-frequency wave fields for which  energy can flow from unstable modes to high harmonic stable modes of AGWs. Thus,  in order to take into account the exchange of energy among AGW spectrum and the flow due to the eventual occurrence of an instability,  we consider the slow variations of both the high-frequency wave $M_k$ and the  low-frequency  current $j_l$, and thereby   replacing $\omega^2$ by $\omega^2 +2i\omega \partial/\partial t$ in   Eq. \eqref{high-freq-dispersion-new} and including the time dependence in $M_k$ for consistency, we obtain \citep{mendonca2015}
\begin{equation}
\left( \frac{\partial}{\partial t}+i \omega\right) M_k(t) + \int \frac{d \bf{q}}{(2\pi)^3} {\cal Q}_k({\bf{q}}) j_q(t)M_{k'}(t)=0,\label{wave-kinetic-eq}
\end{equation}
where $j_q(t)$ is the spatial Fourier components of the nonlinear current ${\bf{j}}_l({\bf{r}},t)$, ${\bf{k'}}={\bf{k}}-{\bf{q}}$ is the new wave vector and the expression ${\cal Q}_k({\bf{q}})$ is given by
\begin{equation}
\begin{split}
{\cal Q}_k({\bf{q}})=&\frac{i c_s^2}{2\omega^2 \rho_0} {\bf{k}} \cdot \left[\left\lbrace (ik_0{\bf e}_z-{\bf{k'}})\right.\right.\\
&\left.\left.-i\frac{2H^2\omega}{c_s^2} \left\lbrace {\bf\Omega}_0 \times (ik_0{\bf e}_z-{\bf{k'}})\right\rbrace \right\rbrace   ({\bf k} \cdot {\bf{e}}_q) \right.\\
&\left. +(ik_0 k'_z-k'^2)(1-i\frac{2H^2\omega}{c_s^2} \Omega_0){\bf{e}}_q\right], \label{coupling-coefficient}
\end{split}
\end{equation}
with  ${\bf{e}}_q\simeq {\bf{j}}_q/|j_q|$ denoting the unit vector. 
\par 
Using the standard Wigner-Moyal formalism and following the work of  \citep{mendonca2015},
we obtain the following wave-kinetic equation for the high frequency perturbations.
\begin{equation}
\begin{split}
\left( \frac{\partial}{\partial t} + {\bf{v}}_{gk} \cdot \nabla\right)W = & \int \frac{d \bf{q}}{(2\pi)^3} {\cal Q}_k({\bf{q}}) J_q(t)\\
&\times[W^- -W^+]\exp(i{\bf{q}}\cdot{\bf{r}}), \label{wave-kinetic-eq-using-wigner}
\end{split}
\end{equation}
where $W\equiv W({\bf{r}},{\bf{k}},t)$ is the Wigner function, given by,
\begin{equation}
W({\bf{r}},{\bf{k}},t)=\int M_h({\bf{r}}-{\bf{s}}/2)M^\ast_h({\bf{r}}-{\bf{s}}/2) e^{i{\bf{k}}\cdot{\bf{s}}} d{\bf{s}},\label{wigner-eq}
\end{equation}
$W^{\pm}=W({\bf{r}},{\bf{k}}\pm {\bf{q}}/2,t)$ and  ${\bf{v}}_{gk}={\partial \omega}/{\partial {\bf{k}}}=c_s^2{\bf{k}}/\omega$ is the group velocity of the high-frequency wave envelope.
 Equation \eqref{wave-kinetic-eq-using-wigner}   describes the evolution of the high-frequency quasi-particles interacting with low-frequency perturbations $j_q(t)$. 
 \par 
Next, in the limit of $|{\bf k}|\gg |{\bf q}|$ (Geometric optics approximation), i.e., if the typical  scale length of low-frequency perturbations with wave vector $q$ is much larger than that of the high-frequency oscillations with wave vector ${\bf k}$,    the difference $[W^- -W^+]$ in Eq. \eqref{wave-kinetic-eq-using-wigner} can be Taylor expanded. Thus, retaining  the lowest order of the Wigner function, Eq. \eqref{wave-kinetic-eq-using-wigner} reduces to the form  of  a kinetic Vlasov equation, given by,
\begin{equation}
\left( \frac{\partial}{\partial t} +{\bf{v}}_{gk} \cdot \nabla +{\bf{F}}_k\cdot \frac{\partial}{\partial {\bf{k}}}\right)W=0. \label{final-wave-kinetic-eq} 
\end{equation}
Here,  $W$ describes the distribution function for the high-frequency atmospheric quasiparticles or phonons,   ${\bf{F}}_k=-\nabla V_k$ is  the  effective nonlinear force acting on the phonons and $V_k({\bf{r}},t)={\cal Q}_k({\bf{q}})J_q(t)\exp(i{\bf{q}}\cdot {\bf{r}})$ is a nonlinear potential associated with the low-frequency perturbations described by the $q$ spectrum.  
 \par 
In what follows,  the evolution equation for the slowly varying current ${\bf{j}}_l(t)$ can be obtained from Eq. \eqref{zakharov-eq-2} in terms of the Wigner function  as
\begin{equation}
\begin{split}
\frac{\partial}{\partial t}{\bf{j}}_l=&-\frac{c_s^4}{4H}\left( 1+\frac{16\Omega_0^2H^2}{c_s^2}\right) \left(1+\frac{2\Omega_0H}{c_s} \right) \\
&\times\left\lbrace \nabla_\perp \frac{\partial}{\partial z}-e_z\nabla_\perp ^2\right\rbrace  \int W({\bf{r}},{\bf{k}},t) \frac{d{\bf{k}}}{(2\pi)^3}. \label{slowly-varing-current-eq}
\end{split}
\end{equation}
Equations \eqref{wave-kinetic-eq-using-wigner} and \eqref{slowly-varing-current-eq} are the desired  wave-kinetic equations equivalent to Eqs. \eqref{zakharov-eq-1} and \eqref{zakharov-eq-2} for   the nonlinear coupling of the high- and low-frequency AGWs that are modified by the Coriolis force. In absence of the latter, one can recover the same equations as in the work of \citep{mendonca2015}.   In order that  Eq. \eqref{slowly-varing-current-eq} includes the linear  internal gravity mode $\Omega=\Omega_q$ [Eq. \eqref{dispersion-low-freq}] in absence of the ponderomotive nonlinearity  for the low-frequency perturbations, i.e., 
\begin{equation}
\Omega\sim \Omega_q\equiv \sqrt{\frac{\omega_g^2 q_\perp^2}{q^2+k_0^2} +4\Omega_0^2}, \label{eq-Omega}
\end{equation}
we replace $\partial/\partial t$ by $\partial/\partial t+i\Omega_q$ in  Eq. \eqref{slowly-varing-current-eq}  and rewrite it as
\begin{equation}
\begin{split}
&\left(\frac{\partial}{\partial t}+i\Omega_q\right){\bf{j}}_l=-\frac{c_s^4}{4H}\left( 1+\frac{16\Omega_0^2H^2}{c_s^2}\right) \left(1+\frac{2\Omega_0H}{c_s} \right) \\
&\times\left\lbrace \nabla_\perp \frac{\partial}{\partial z}-e_z\nabla_\perp ^2\right\rbrace  \int W({\bf{r}},{\bf{k}},t) \frac{d{\bf{k}}}{(2\pi)^3}. \label{eq-jl}
\end{split}
\end{equation}

\section{Nonlinear dispersion relation}\label{sec-stabi}
In this section, we study the stability of large scale ($|{\bf k}|\gg|{\bf q}|$) low-frequency  perturbations by deriving an approximate nonlinear dispersion relation. To this end, we assume a low-frequency perturbation associated with the current  of the form ${\bf{j}}_l({\bf{r}},t)={\bf{j}}_q\exp(i{\bf{q}}\cdot{\bf{r}}-i\Omega t)$  and  that this mode approximately satisfies the low-frequency dispersion equation \eqref{eq-Omega}. 
 Thus, from Eq. \eqref{eq-jl} we obtain \citep{mendonca2015}  
\begin{equation}
\begin{split}
\left(\Omega-\Omega_q \right)&{\bf{j}}_q =i\frac{c_s^4}{4H}\left( 1+\frac{16\Omega_0^2H^2}{c_s^2}\right) \left(1+\frac{2\Omega_0H}{c_s} \right) \\
&\times\left\lbrace {\bf{q}}_\perp q_z-{\bf{e}}_z q_\perp ^2 \right\rbrace  \int W_q({\bf{r}},{\bf{k}}) \frac{d{\bf{k}}}{(2\pi)^3}, \label{slowly-varing-current-final-eq}
\end{split}
\end{equation}
where $W_q({\bf{r}},{\bf{k}})$ denotes the modulation of the quasi-distribution function $W({\bf{r}},{\bf{k}},t)$ under the low-frequency plane wave perturbation.    The value of $W_q$ can be obtained by linearizing the wave kinetic equation  \eqref{wave-kinetic-eq-using-wigner} for high-frequency waves as
\begin{equation}
W_q=Q_k({\bf{q}}) \frac{[W_0^- -W_0^+]}{\Omega -{\bf{q}}\cdot{\bf{v}}_k}, \label{linearize-wq-eq}
\end{equation}
where $W_0$ is the is the unperturbed quasi-particle distribution function and $W_0^{\pm}\equiv W_0({\bf{k}}\pm {\bf{q}}/2)$. Thus, using Eq. \eqref{linearize-wq-eq}, we obtain from Eq. \eqref{slowly-varing-current-final-eq} the following nonlinear dispersion relation for a low-frequency wave mode with  frequency $\Omega$ and  wave vector $\bf{q}$ that is driven by the arbitrary spectrum of high-frequency perturbations.
\begin{equation}
\begin{split}
&1-\frac{\Omega_q}{\Omega}-i\frac{c_s^4}{4H\Omega}\left( 1+\frac{16\Omega_0^2H^2}{c_s^2}\right) \left(1+\frac{2\Omega_0H}{c_s} \right)\\
&\times\left\lbrace {\bf{q}}_\perp q_z-{\bf{e}}_z q_\perp ^2 \right\rbrace \cdot {\bf{e}}_q \int Q_k({\bf{q}}) \frac{[W_0^- -W_0^+]}{\Omega -{\bf{q}}\cdot{\bf{v}}_k} \frac{d{\bf{k}}}{(2\pi)^3}=0, \label{dispersion-low-frequency-final}
\end{split}
\end{equation}
  Here, ${\bf{e}}_q=[({\bf{q}}_\perp)/q_\perp -{\bf{e}}_z q_\perp]/q$,  indicating that the nonlinear current  and the wave   vector are perpendicular to each other.  From Eq. \eqref{dispersion-low-frequency-final}, we note that the dispersion relation is significantly modified by the effects of the Coriolis force. Furthermore, the Landau resonance occurs when the group velocity $v_{gk}$ of the high-frequency quasi-particles approaches the phase velocity $\Omega/q$ of the low-frequency perturbations. It is also noticed that the resonant velocity of the quasi-particles is up-shifted by a quantity $\propto\Omega_0^2$ as the   phase velocity $\Omega/q$   is increased   and $v_{gk}$  remains unaltered by the influence of the Coriolis force [\textit{cf}. Eqs. \eqref{dispersion-high-freq}, \eqref{dispersion-low-freq}].   It follows that the wave-kinetic approach provides an alternative mechanism for the transfer of wave energy in the interactions of  high- and low-frequency modes of AGWs in the  atmosphere. 

\section{Conclusion} \label{sec-conclu}
We have studied the influence of the Coriolis force on the nonlinear interactions of high- and low-frequency branches of  AGWs in the atmosphere.   Starting from a set of fluid equations modified by the Coriolis force the two linear dispersion branches are obtained in two different limits, namely  $\omega\gg\omega_a$ (high-frequency) and $\omega\ll\omega_g$ (low-frequency). While the high-frequency acoustic mode remains unaltered, the low-frequency internal mode gets modified by the Earth's uniform angular velocity. Following the work of \citep{mendonca2015}, the nonlinear coupling of these two modes are described by two-equivalent approaches: the Zakharov approach and the wave-kinetic approach. In the former, the ponderomotive nonlinearity, associated with the high-frequency fields, gets enhanced by the effects of the Coriolis force. This may eventually lead to an increase of the soliton amplitude to be formed due to the nonlinear interactions or a development of the chaotic aspects of the system. As a result, the energy transfer between the high- and low-frequency modes may become faster the larger is the possibility of the emergence of atmospheric turbulence.    On the other hand, an approximate  nonlinear dispersion relation for the low-frequency IGWs  is obtained from the wave-kinetic equations in presence of an arbitrary spectrum of high-frequency atmospheric phonons, which indicates that the Landau resonance condition is modified by the Coriolis force, i.e., the resonant velocity  of  high-frequency quasi-particles gets up-shifted by a quantity $\propto\Omega_0^2$.
\par 
 It is worthwhile to mention that the coupled high- and low-frequency modes of AGWs [also known as the inertio-gravity waves \citep{kaladze2007}] that are generated by the combined influence of the gravitational force and the Coriolis force can propagate in the regions  of lower, middle or upper Earth's atmosphere (e.g., ionospheric  D, E   or F layers). Such waves, while interacting   with other waves or atmospheric charged particles, can break and produce different kinds of disturbances \citep{snively2003,chen2015}. In presence of the geomagnetic field they may be dissipated [due to Pedersen conductivity \citep{kaladze2008}]  which may, in turn, generate jet streams and change the heat balance in the upper atmosphere \citep{fritts2006,karpov2017}. Furthermore, the AGWs reaching the Earth's ionosphere   can influence the motion of  plasma particles and hence the radio wave transmission. 
 \par 
   It is to be noted that  the Zakharov approach is more adequate than the wave-kinetic approach  for the description of solitons where the formation of electrostatic or electromagnetic wave envelope is highly correlated with the density   depletion \citep{banerjee2010}.  On the other hand, the wave-kinetic approach describes the energy exchange between low-frequency waves and high-frequency quasi-particles due to resonance with the group velocity \citep{mendonca2015}.  
\par 
To conclude, at high altitudes, the motion of atmospheric charged particles may be significantly influenced by the    Ampere force $({\bf j}\times {\bf B})$ force. So, the inclusion of this force in the fluid dynamics may introduce a new physical effect to the nonlinear coupling of AGWs and IGWs. However, such an investigation is left for a future project.     
%%%%%%%%%%%%%%%%%%%%%%%%%%%%%%%%%%% 
\section*{Acknowledgement} {The authors wish to thank Professor Lennart Stenflo of Link{\"o}ping University, Sweden for his valuable suggestions.    D. Chatterjee acknowledges support from Science and Engineering Research Board (SERB)  for a national postdoctoral fellowship (NPDF)  with sanction order no. PDF/2020/002209 dated 31 Dec 2020. A. P. Misra thanks   SERB  for support through a project with sanction order no. CRG/2018/004475. }
%%%%%%%%%%%%%%%%%%%%%%%%%%%%%%%%%%%%%%%%%%%%%%%%%%%%%%%%%%
 
%\section{Bibliography}
%\printcredits

%% Loading bibliography style file
%\bibliographystyle{model1-num-names}
\bibliographystyle{cas-model2-names}

% Loading bibliography database
\bibliography{ref}

%% Loading bibliography style file
%\bibliographystyle{model1-num-names}
%\bibliographystyle{cas-model2-names}

% Loading bibliography database
%\bibliography{cas-refs}

%\vskip3pt

%\bio{}
%Author biography without author photo.
%Author biography. Author biography. Author biography.
%\endbio
%
%\bio{figs/pic1}
%Author biography with author photo.
%Author biography. Author biography. Author biography.
%\endbio
%
%\bio{figs/pic1}
%Author biography with author photo.
%Author biography. Author biography. Author biography.
%\endbio

\end{document}